\documentclass[
showpacs,
floatfix,
aps,
prb,
twocolumn,
superscriptaddress,
amssymb,
]{revtex4-1}

\usepackage{amssymb}
\usepackage{latexsym}
\usepackage{graphicx}
\usepackage{amsmath}
\usepackage{graphicx}
\usepackage{dcolumn}
\usepackage{amsfonts}
\usepackage{bm}
\usepackage{epsfig}

\newcommand{\be}{\begin{equation}}
\newcommand{\ee}{\end{equation}}
\newcommand{\bea}{\begin{eqnarray}}
\newcommand{\eea}{\end{eqnarray}}

\def\xt{\mathcal{X}_{2}}
\def\m{m^\star}

\def\pac{\delta}

\def\oc{\omega_{\mbox{\scriptsize {c}}}}
\def\op{\omega_0}

\def\mb{{m^\star_{\rm b}}}

\def\ef{\varepsilon_F}

\newcommand{\req}[1]{Eq.\,(\ref{#1})}
\newcommand{\rEq}[1]{Equation\,(\ref{#1})}

\newcommand{\rfig}[1]{Fig.\,\ref{#1}}
\newcommand{\rFig}[1]{Figure \,\ref{#1}}
\newcommand{\rfigs}[2]{Figs.\,\ref{#1} and \ref{#2}}

\newcommand{\rref}[1]{Ref.\,\onlinecite{#1}}
\newcommand{\rrefs}[2]{Refs.\,\onlinecite{#1} and \onlinecite{#2}}
\newcommand{\rrefss}[3]{Refs.\,\onlinecite{#1},\,\onlinecite{#2} and \onlinecite{#3}}

\def\moba{$\mu_A \approx 1.3 \times 10^7$ cm$^2$/Vs}
\def\dena{$n_A \approx 2.7 \times 10^{11}$ cm$^{-2}$}
\def\mobb{$\mu_B \approx 1.1 \times 10^7$ cm$^2$/Vs}
\def\denb{$n_B \approx 3.2 \times 10^{11}$ cm$^{-2}$}
\def\ne{n_e}

\begin{document}
\title{
Evidence for effective mass reduction in GaAs/AlGaAs quantum wells
}
\author{A.\,T. Hatke}
\affiliation{School of Physics and Astronomy, University of Minnesota, Minneapolis, Minnesota 55455, USA}

\author{M.\,A. Zudov}
\email[Corresponding author: ]{zudov@physics.umn.edu}
\affiliation{School of Physics and Astronomy, University of Minnesota, Minneapolis, Minnesota 55455, USA}

\author{J.\,D. Watson}
\affiliation{Department of Physics, Purdue University, West Lafayette, Indiana 47907, USA}
\affiliation{Birck Nanotechnology Center, School of Materials Engineering and School of Electrical and Computer Engineering, Purdue University, West Lafayette, Indiana 47907, USA}

\author{M.\,J. Manfra}
\affiliation{Department of Physics, Purdue University, West Lafayette, Indiana 47907, USA}
\affiliation{Birck Nanotechnology Center, School of Materials Engineering and School of Electrical and Computer Engineering, Purdue University, West Lafayette, Indiana 47907, USA}
\received{February 15, 2012, revised manuscript received April 6 2013, published April 26 2013}

\author{L.\,N. Pfeiffer}
\affiliation{Department of Electrical Engineering, Princeton University, Princeton, New Jersey 08544, USA}

\author{K.\,W. West}
\affiliation{Department of Electrical Engineering, Princeton University, Princeton, New Jersey 08544, USA}

\begin{abstract}
We have performed microwave photoresistance measurements in high mobility GaAs/AlGaAs quantum wells and investigated the value of the effective mass. 
Surprisingly, the effective mass, obtained from the period of microwave-induced resistance oscillations, is found to be about $12$\,\% {\em lower} than the band mass in GaAs, $\mb$.
This finding provides strong evidence for electron-electron interactions which can be probed by microwave photoresistance in very high Landau levels.
In contrast, the measured magnetoplasmon dispersion revealed an effective mass which is close to $\mb$, in accord with previous studies.
\end{abstract}
\maketitle

The most frequently quoted value of the effective mass $\m$ in GaAs/AlGaAs-based two-dimensional electron systems (2DES) is the value of the band mass of bulk GaAs, $\mb=0.067\,m_0$ ($m_0$ is the free electron mass) \cite{sze:1981}.
One of the oldest and still frequently employed experimental methods to obtain $\m$ is based on Shubnikov-de Haas oscillations (SdHOs).\cite{shoenberg:1984,pudalov:2011}
Being a result of Landau quantization in a magnetic field $B$, SdHOs are controlled by the filling factor, $\nu = 2\ef/\hbar\oc$, where $\ef = \pi \hbar^2 \ne/\m$ is the Fermi energy,  $\ne$ is the carrier density, and $\hbar\oc=eB/\m$ is the cyclotron energy.
Since $\m$ does not enter the filling factor, it cannot be obtained from the oscillation period but, instead, one has to analyze the temperature damping of the SdHO amplitude.

The SdHO approach applied to 2DES with $\ne \gtrsim 10^{11}$ cm$^{-2}$ usually yields $\m$ values which are close to, or somewhat higher than, $\mb$.\citep{smrcka:1995,hang:2007}
However, there exist studies \citep{coleridge:1996,tan:2005} which report values significantly ($\simeq$ 10 \%) lower than $\mb$.
The disagreement in obtained mass values can, at least in part, be accounted for by a relatively low accuracy of the SdHO approach.\citep{note:8} 
There also exist other factors which might affect extracted $\m$, even when the procedure seems to work properly.\citep{hayne:1992,hayne:1997,coleridge:1996,coleridge:1991}
According to \rref{tan:2005}, the lower values of $\m$ might very well be a signal of electron-electron interactions which, in contrast to the case of dilute 2DES, can actually {\em reduce} the effective mass at intermediate densities.\citep{smith:1992,kwon:1994,zhang:2005b,asgari:2005,asgari:2006,drummond:2009}
Therefore, it is both interesting and important to revisit the issue of low effective mass values using alternative experimental probes, which we do in this paper.

In addition to SdHOs, several other types of magnetoresistance oscillations are known to occur in high mobility 2DES.\citep{zudov:2001a,zudov:2001b,yang:2002,zhang:2007c,zhang:2008,hatke:2008a,hatke:2008b,khodas:2010}
Unlike the filling factor entering SdHOs, the parameters controlling these oscillations {\em do depend} on $\m$, thus making it available directly from the oscillation period.
In what follows, we briefly discuss one such oscillation type, microwave-induced resistance oscillations (MIROs),\cite{zudov:2001a} whose period can be  measured with high precision.

MIROs appear in magnetoresistivity when a 2DES is irradiated by microwaves.
Being a result of electron transitions between Landau levels owing to photon absorption, MIROs are controlled by $\omega/\oc$, where $\omega = 2\pi f$ is the radiation frequency.
It is well established both theoretically \citep{durst:2003,vavilov:2004,dmitriev:2005,dmitriev:2009b,dmitriev:2012} and experimentally,\citep{mani:2002,zudov:2004,mani:2004e,hatke:2009a} that MIROs can be described by $-\sin(2\pi\omega/\oc)$, provided that $2\pi\omega/\oc \gg 1$ and that the microwave power is not too high.\citep{hatke:2011e}
As a result, the higher order ($i = 3,4,...$) MIRO maxima are \emph{accurately}\citep{hatke:2011f} described by
\be
\omega = \frac {e}{\m}B_i (i - \pac) \,,
\label{eq.miro}
\ee
where 
$B_i$ is the magnetic field of the $i$th maximum and $\pac \approx 1/4$.\citep{note:11}
Once the value of $\pac$ is verified experimentally, one can obtain $\m$ using, e.g., the dispersion of the $i$th MIRO maximum, $f(B_i)$.\citep{note:4}
Equivalently, the mass can be obtained directly from the oscillation period at a given $\omega$, e.g., from the dependence of $i$ on $B_i$, $i = \omega \m/e B_i +\pac$.

In this paper we investigate the effective mass in very high mobility GaAs/AlGaAs quantum wells using microwave photoresistance measurements performed over a wide frequency range from 100 GHz to 175 GHz.
Remarkably, the effective mass extracted from MIROs is found to be considerably lower than the band mass value.
More specifically, MIROs are found to be well described by \req{eq.miro} with the effective mass $\m\approx 0.059 \,m_0$ at \emph{all} frequencies studied.
These findings provide strong evidence for electron-electron interactions which can be probed by microwave photoresistance in very high Landau levels.
In contrast, the measured dispersion of the magnetoplasmon resonance (MPR) reveals $\m \approx \mb$, in agreement with previous studies.

Our sample A (sample B) is a lithographically defined Hall bar of width $w_A=50$ $\mu$m ($w_{\rm B}=200$ $\mu$m) fabricated from a 300 \AA-wide GaAs/Al$_{0.24}$Ga$_{0.76}$As quantum well grown by molecular beam epitaxy at Purdue (Princeton).
The low-temperature density and  mobility of sample A (sample B) were \dena~(\denb) and \moba~(\mobb), respectively. 
Microwave radiation, generated by a backward wave oscillator, was delivered to the sample placed in a $^{3}$He cryostat via a 1/4-in.- (6.35-mm)-diameter light pipe. 
The resistivity $\rho_\omega$ was measured under continuous microwave irradiation using a standard low-frequency lock-in technique.

\begin{figure}[t]
\includegraphics{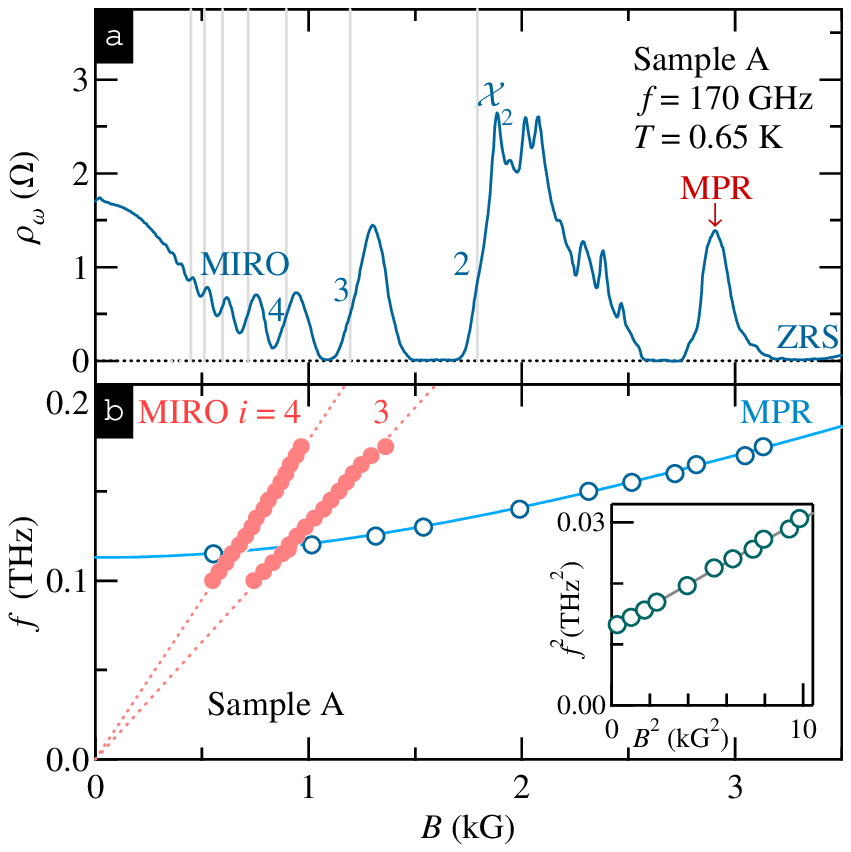}
\vspace{-0.1 in}
\caption{(Color online)
(a) Magnetoresistivity $\rho_\omega(B)$ measured at $T = 0.65$ K in sample A irradiated with microwaves of $f = 170$ GHz.
The vertical lines (marked by $i$) are drawn at the harmonics of the cyclotron resonance, $\omega/\oc=i$, calculated using $\m = 0.0590 m_0$.
(b) Dispersions $f(B)$ of the MIRO maxima for $i=3,4$ (solid circles) and of the MPR peak (open circles).
The dotted lines are fits to the data, $f=(i-1/4)eB_i/2\pi \m$, with $i=3,4$.
The solid curve is calculated from $f = \sqrt{f_0^2+(eB/2\pi \m)^2}$, using $f_0=112.5$ GHz and $\m = 0.066\,m_0$.
The inset shows the MPR dispersion as $f^2$ vs $B^2$ (circles) and a linear fit (solid line), $f^2 = f_0^2+(eB/2\pi \m)^2$, which yields $f_0$ and $\m$ quoted above.  
}
\vspace{-0.15 in}
\label{fig1}
\end{figure}

In \rfig{fig1}(a) we present the magnetoresistivity $\rho_\omega(B)$ measured at $T = 0.65$ K in sample A under microwave irradiation of frequency $f = 170$ GHz.
The data reveal a giant negative magnetoresistance effect,\citep{bockhorn:2011,hatke:2012j} pronounced MIROs, zero-resistance states,\citep{mani:2002,zudov:2003,andreev:2003,smet:2005,zudov:2006a,zudov:2006b,dorozhkin:2011} and a strong peak (marked by ``MPR'') which corresponds to the lowest mode of the dimensional magnetoplasmon resonance.\citep{vasiliadou:1993,kukushkin:2006b,hatke:2012h}
Finally, we notice a series of fast oscillations superimposed on the second MIRO maximum.
The origin of these oscillations is unknown at this point, but the peak closest to the second harmonic of the cyclotron resonance (marked by ``$\xt$'') looks similar to the recently discovered radiation-induced $\xt$ peak.\citep{dai:2010,dai:2011,hatke:2011b,hatke:2011c,hatke:2011f}
We will return to this peak when we discuss our results in sample B. 

One can accurately determine the effective mass entering \req{eq.miro} by trial and error, namely, by adjusting $\m$ until {\em each} calculated cyclotron resonance harmonic falls symmetrically between maximum and minimum of the same order.
Remarkably, such a procedure applied to the data in \rfig{fig1}(a) results in $\m = 0.059\,m_0$, used to calculate the positions of vertical lines (marked by $i$) drawn at $\omega/\oc=i=2,3,4,\dots\,$.
The obtained value is considerably ($\approx 12$ \%) {\em lower} than $\mb=0.067\,m_0$ and its confirmation warrants further investigation. 

To this end, and to confirm that the strong peak in \rfig{fig1}(a) is due to MPR, we have repeated our measurements at a variety of microwave frequencies, from 100 to 175 GHz.
From these data we have then extracted the magnetic field positions of the MIRO maxima and of the MPR peak for all frequencies studied.
Our findings are presented in \rfig{fig1}(b) showing microwave frequency $f$ as a function of $B$ corresponding to $i=3,4$ MIRO maxima (solid circles) and to the MPR peak (open circles).
It is clear that the MIRO maxima follow the expected linear dispersion relation, which extrapolates to the origin, as expected from \req{eq.miro}. 
By fitting the data (dotted lines) with \req{eq.miro}, $f=(i-1/4)eB_i/2\pi \m$, we obtain $\m=0.0586\,m_0$ and $\m=0.0587\,m_0$ for $i=3$ and $i=4$, respectively.
Since the obtained values are both very close to each other, we conclude that the effective mass $\m \approx 0.059\,m_0$ accurately describes MIRO in sample A.

On the other hand, the MPR peak follows a dispersion [cf. open circles in \rfig{fig1}(b)] characteristic of a magnetoplasmon resonance,\citep{chaplik:1972}
\be
\omega^2= \oc^2+\op^2\,,
\label{eq.mpr}
\ee
where $\op$ is the frequency of the lowest mode of standing plasmon oscillation.
As shown in the inset, $f^2$ is a linear function of $B^2$, in agreement with \req{eq.mpr}.
From the slope of the fit to the data with $f^2 = f_0^2+(eB/2\pi \m)^2$ (cf. solid line in the inset) we obtain $\m \approx 0.066 m \approx \mb$.
We also notice that previous MPR experiments obtained $\m$ values ranging from 0.067 to 0.071.\citep{vasiliadou:1993,yang:2006,kukushkin:2006b,muravev:2007,dorozhkin:2007}

Using $\op \approx 0.85\sqrt{\pi e^2\ne/2\varepsilon_0\bar\varepsilon\m w}$,\citep{stern:1967,vasiliadou:1993,mikhailov:2004,mikhailov:2005,note:9} where $\m=0.066\,m_0$, $\varepsilon_0$ is the permittivity of vacuum, and $\bar\varepsilon = 6.9$ is the average dielectric constant of GaAs (12.8) and free space (1), we estimate $f_0=\op/2\pi \approx 105$ GHz.
This value is in good agreement with $f_0 \approx 112$ GHz obtained from the value of the fit at $B^2=0$.
The MPR dispersion $f(B)$ [cf.\,solid curve in \rfig{fig1}(b)], calculated using \req{eq.mpr} and extracted $f_0$ and $\m$, shows excellent agreement with our experimental data. 
We thus conclude that the peak marked by ``MPR'' in \rfig{fig1}(a) originates from the fundamental MPR mode.\citep{vasiliadou:1993,kukushkin:2006b,hatke:2012h}

\begin{figure}[t]
\includegraphics{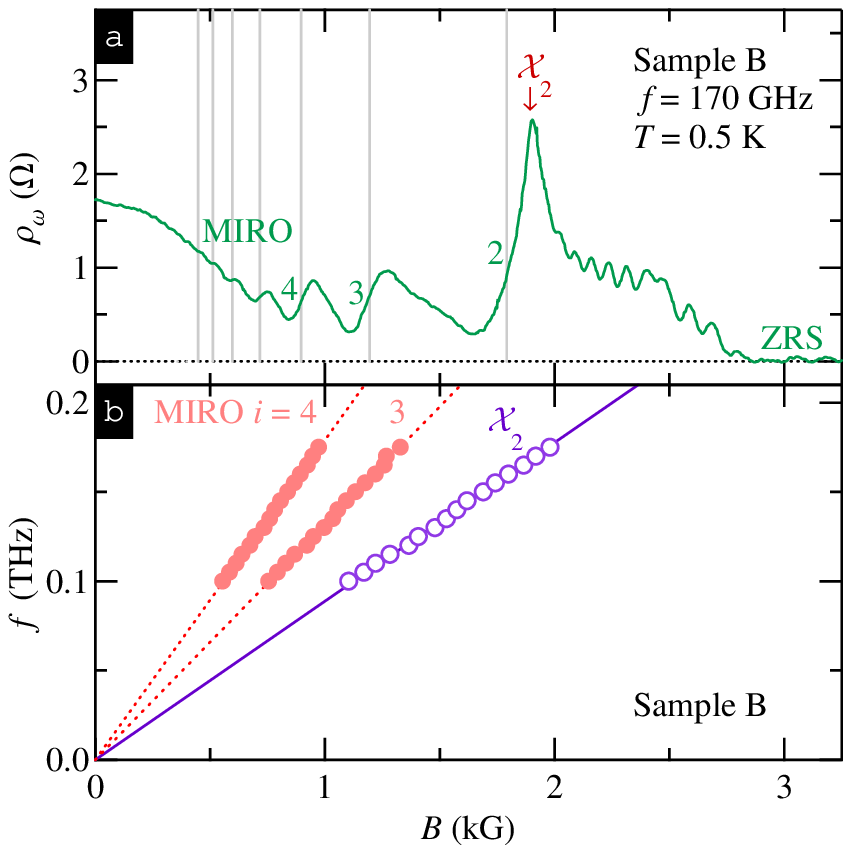}
\vspace{-0.1 in}
\caption{(Color online)
(a) Magnetoresistivity $\rho_\omega(B)$ measured at $T = 0.5$ K in sample B irradiated with microwaves of $f = 170$ GHz.
The vertical lines (marked by $i$) are drawn at the harmonics of the cyclotron resonance, $\omega/\oc=i$, calculated using $\m = 0.0590\,m_0$.
(b) Dispersions $f(B)$ of the MIRO maxima for $i=3,4$ (solid circles) and of the $\xt$ peak (open circles).
The dotted lines are fits to the data, $f=(i-1/4)eB_i/2\pi \m$, with $i=3,4$.
The fit to the $\xt$ dispersion (cf.\,solid line), $f = eB/\pi \m$, yields $\m = 0.0630\,m_0$.
}
\vspace{-0.15 in}
\label{fig2}
\end{figure}

The main conclusion of our study on sample A is that the effective mass obtained from MIROs is significantly lower than both the mass entering the magnetoplasmon resonance and the band mass in GaAs.
To confirm this finding we have performed similar measurements on sample B.
\rFig{fig2}(a) shows $\rho_\omega(B)$ measured at $T = 0.5$ K in sample B under microwave irradiation of $f = 170$ GHz.
Following the procedure of trial and error, we again find that aligning MIROs with the harmonics of cyclotron resonance (cf. vertical lines) calls for a low value of the effective mass, $\m = 0.059\,m_0$.
By repeating the measurements at different $f$ from 100 to 175 GHz, we have obtained the dispersion relations for the $i=3$ and $i=4$ MIRO maxima, which are shown in \rfig{fig2}(b) as solid circles.
The linear fits with $f=(i-1/4)eB_i/2\pi \m$ generate $\m = 0.0584\,m_0$ and $\m=0.0586\,m_0$ for $i=3$ and $i=4$, respectively.
We thus again find a considerably reduced effective mass value which nearly matches our result in sample A.

Close examination of \rfig{fig2}(a) reveals that the photoresistance maximum near the second harmonic of the cyclotron resonance is considerably higher and sharper than all other maxima.
We attribute this maximum to the $\xt$ peak recently discovered in high mobility 2DES.\citep{dai:2010,dai:2011,hatke:2011b,hatke:2011c,hatke:2011f}
While the origin of the $\xt$ peak remains unknown, its large amplitude\citep{dai:2010,hatke:2011b} and distinct responses to dc\citep{hatke:2011c} and to in-plane magnetic\citep{dai:2011} fields strongly support the notion that the $\xt$ peak and MIROs are two different phenomena.
However, there exists a controversy regarding its exact position. 
More specifically, \rrefs{dai:2010}{dai:2011} concluded that the $\xt$ peak occurs {\em exactly} at the second harmonic of the cyclotron resonance, $\omega/\oc = 2$.
However, \rrefss{hatke:2011f}{hatke:2011b}{hatke:2011c} found that the peak occurs at somewhat higher $B$ than the second harmonic.
This apparent controversy can be resolved by noticing that the above conclusions were made based on different approaches. 
While \rref{dai:2010} has determined the $\xt$ peak position from the cyclotron resonance measured in absorption, \rrefss{hatke:2011f}{hatke:2011b}{hatke:2011c} used MIROs as a reference.
Indeed, using the latter approach we find that the $\xt$ peak occurs at a magnetic field somewhat higher than the second harmonic, as in previous studies.\citep{hatke:2011b,hatke:2011c,hatke:2011f}

On the other hand, we have just established that the MIRO effective mass is significantly lower than the mass entering the MPR, which is closely related to the cyclotron resonance.
Therefore it is interesting to examine the effective mass obtained from the $\xt$ peak, assuming that it appears {\em exactly} at the second harmonic of the cyclotron resonance, as found in \rrefs{dai:2010}{dai:2011}.
As shown in \rfig{fig2}(b) by open circles, the $\xt$ peak follows a linear dispersion relation extrapolating through the origin.
A linear fit with $f = eB/\pi\m$, shown by the solid line, generates $\m = 0.063\,m_0$,\citep{note:10} which is noticeably higher (lower) than the MIRO (MPR) mass.

\begin{figure}[t]
\includegraphics{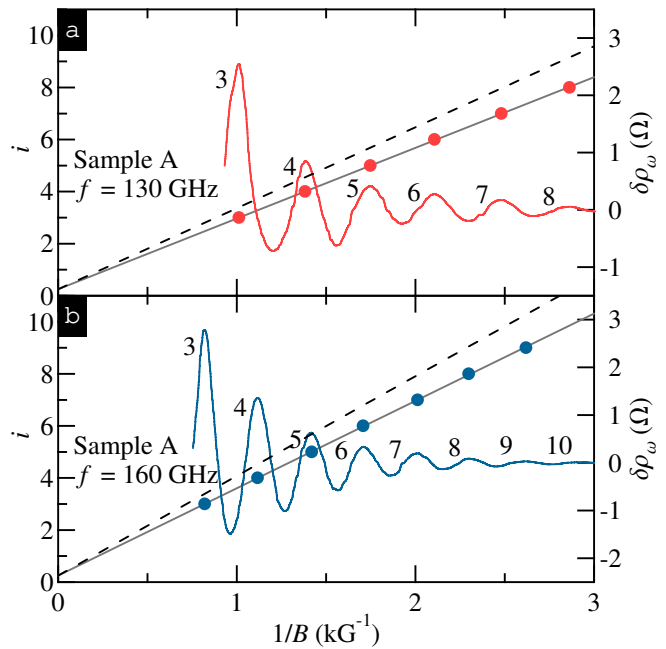}
\vspace{-0.1 in}
\caption{(Color online)
Microwave photoresistivity $\delta\rho_\omega$ (right axis, solid curve) and the order of the MIRO maxima $i$ (left axis, circles) vs. $1/B$ measured in sample A at (a) $f=130$ GHz and (b) $f=160$ GHz. 
Fits to the data (solid lines) with $i = 2 \pi f \m/e B + \pac$ yield $\pac \approx 0.25$ and $\m \approx 0.0585\,m_0$ ($\m \approx 0.0587\,m_0$) for $f=130$ GHz ($f=160$ GHz). 
Dashed lines are calculated using \req{eq.miro} and $\m=\mb=0.067\,m_0$.
}
\vspace{-0.15 in}
\label{fig3}
\end{figure}

As mentioned above, one can also obtain $\m$ directly from the MIRO period.
This method is based on scaling of multiple oscillations and does not \emph{a priori} assume $\pac = 1/4$. 
To illustrate this approach, we present on the right axis of \rfig{fig3} microwave photoresistivity $\delta\rho_\omega = \rho_\omega - \rho$ as a function of $1/B$ measured in sample A at (a) $f=130$ GHz and (b) $f=160$ GHz. 
Both data sets exhibit multiple oscillations whose period scales with $1/\m f$.
To extract $\m$ from the data, we plot the order of the MIRO maxima $i$ (circles, left axis) as a function of $1/B$ for both frequencies and observe expected linear dependence.
From the slope of the linear fits to the data (solid lines), $i = 2 \pi f \m/e B + \pac$, we find $\m \approx 0.0585\,m_0$ ($\m \approx 0.0587\,m_0$) for $f=130$ GHz ($f=160$ GHz).\citep{note:12}
These values are in excellent agreement with the $\m$ values found from the dispersions of the $i=3,4$ MIRO maxima (cf. \rfigs{fig1}{fig2}).
In addition, we find that both fits intercept the vertical axis at $\pac \approx 0.25$, in agreement with \req{eq.miro}, confirming the equivalence of two approaches.
Finally, to illustrate that our data cannot be described by the band mass, we include dashed lines which are calculated using $i = 2 \pi f \mb/e B + 0.25$.

\begin{figure}[t]
\includegraphics{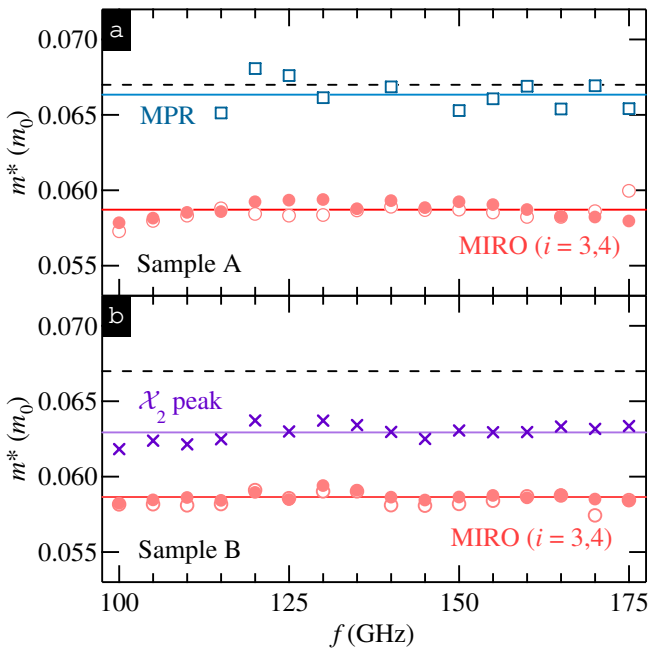}
\vspace{-0.1 in}
\caption{(Color online)
(a) $\m$ obtained from the MIRO maxima for $i=3$ (open circles), $i=4$ (filled circles) and the MPR peak (squares) vs $f$ measured in sample A.
Solid lines represent averages for the $i=4$ MIRO maxima, $\m = 0.0587\,m_0$ and for the MPR peak, $\m = 0.0664\,m_0$, respectively.
(b) $\m$ obtained from the MIRO maxima for $i=3$ (open circles), $i=4$ (filled circles) and from the $\xt$ peak (crosses) vs $f$ measured in sample B.
Solid lines represent averages for the $i=4$ MIRO maxima, $\m = 0.0586\,m_0$, and for the $\xt$ peak $\m = 0.0629\,m_0$ (see text), respectively.
In both plots, the dashed lines represent $\mb = 0.067\,m_0$.
}
\vspace{-0.15 in}
\label{fig4}
\end{figure}

We summarize our findings in \rfig{fig4}, showing effective mass values, obtained from the dispersion relations of different phenomena, as a function of microwave frequency.
More specifically, $\m$ obtained from the MIRO maxima for $i=3$ (open circles) and $i=4$ (solid circles)  measured in samples A and B are shown in \rfig{fig4}(a) and \rfig{fig4}(b), respectively.
In addition, \rfig{fig4}(a) shows $\m$ obtained from the MPR (squares), while \rfig{fig4}(b) shows $\m$ from the $\xt$ peak, assuming that it occurs at the second cyclotron resonance harmonic. 
Solid horizontal lines represent the averages of the measured values (see figure caption) and dashed horizontal lines are drawn at $\mb = 0.067\,m_0$.
\rFig{fig4} further confirms that the masses extracted from the fits in \rfig{fig1}(b) and \rfig{fig2}(b) accurately describe our experimental data over the entire range of frequencies studied.

In summary, we have investigated microwave photoresistance in very high mobility GaAs/AlGaAs quantum wells over a wide range of microwave frequencies.
The analysis of the period of microwave-induced resistance oscillations reveals an effective mass $\m \approx 0.059\,m_0$, which is considerably lower than the GaAs band mass $\mb=0.067\,m_0$.
These findings provide strong evidence for electron-electron interactions in very high Landau levels and for sensitivity of MIROs to these interactions.
On the other hand, the measured dispersion of the magnetoplasmon resonance is best described by $\m \approx \mb$.
It would be interesting to examine if the low value of the effective mass is confirmed in studies of other nonlinear phenomena, such as Hall-field induced resistance oscillations.\citep{yang:2002,zhang:2007,hatke:2009m,hatke:2011n}

We thank M. Dyakonov and B. Shklovskii for discussions and J. Jaroszynski, J. Krzystek, G. Jones, T. Murphy, and D. Smirnov for technical assistance.
This work was supported by the US Department of Energy, Office of Basic Energy Sciences, under Grant Nos. DE-SC002567 (Minnesota) and DE-SC0006671 (Purdue). 
A portion of this work was performed at the National High Magnetic Field Laboratory (NHMFL), which is supported by NSF Cooperative Agreement No. DMR-0654118, by the State of Florida, and by the DOE.
The work at Princeton was partially funded by the Gordon and Betty Moore Foundation and the NSF MRSEC Program through the Princeton Center for Complex Materials (DMR-0819860).


\begin{thebibliography}{55}
\bibitem[{not({\natexlab{c}})}]{note:4}
\bibinfo{note}{Alternatively, $\m$ can be obtained from (a) the MIRO minima, described by $\omega = \oc(i+1/4)$, (b) the zero-response nodes, $\omega=\oc \cdot i$, where microwave photoresistance vanishes, or (c) directly from the oscillation period, which can be found from the positions of, e.g., $i$th and $(i+1)$th maxima.}
\bibitem[{not({\natexlab{d}})}]{note:9}
\bibinfo{note}{\rEq{eq.mpr} is valid when the retardation effects can be
  ignored, i.e. when $\alpha = \sqrt{e^2n_ew/2\pi\varepsilon_0\m c^2} < 1$
  (\rrefs{mikhailov:2004}{mikhailov:2005}). In sample A, we estimate $\alpha
  \simeq 0.15$.}
\bibitem[{not({\natexlab{f}})}]{note:12}
\bibinfo{note}{Similar $\m$ values have been obtained from the MIRO minima and from the zero-response nodes (not shown).}
\end{thebibliography}
\end{document}